# Photo-thermoelectric detection of cyclotron resonance in asymmetrically carrier-doped graphene two-terminal device

Kei Kinoshita[1], Rai Moriya[1,*], Miho Arai[1], Satoru Masubuchi[1], Kenji Watanabe[2], Takashi Taniguchi[2], and Tomoki Machida[1,*]

[1] *Institute of Industrial Science, University of Tokyo, 4-6-1 Komaba, Meguro, Tokyo 153-8505, Japan*

[2] *National Institute for Materials Science, 1-1 Namiki, Tsukuba 305-0044, Japan*

**Graphene is known to show a significant photo-thermoelectric effect that can exceed its photovoltaic contribution. Here, by utilizing this effect, we demonstrate a photovoltage measurement of cyclotron resonance in a double-back-gated h-BN/graphene/h-BN two-terminal device. A graphite local bottom gate was fabricated in addition to a p-doped Si global back gate. By tuning the two gate voltages, an in-plane graphene junction having an asymmetric carrier-doping profile was created. With the help of this asymmetric structure, the photo-thermoelectric voltage generated in the vicinity of the metal-electrode/graphene junction was detected. At a low temperature and in the presence of a magnetic field, a photo-induced voltage was measured under the irradiation of an infrared laser ($\lambda$ = 9.28–10.61 μm). We observed a strong enhancement of the photovoltage signal under the cyclotron resonance condition, at which the energy of excitation coincides with a transition between Landau levels. These results highlight the possibility of using the photo-thermoelectric effect in graphene for THz photo-detection.**

*E-mail: moriyar@iis.u-tokyo.ac.jp; tmachida@iis.u-tokyo.ac.jp

Graphene exhibits an unconventional Landau quantization owing to its unique band structure. Thus, it has been studied extensively to reveal quantum Hall physics [1,2]. The energy of a Landau level of graphene changes in proportion to $\sqrt{B}$, where $B$ denotes the magnetic field perpendicular to graphene's plane, and the spacings between different Landau levels are unequal. The optical selection rule between the Landau levels of graphene is $\Delta|N| =\pm 1$, where $N$ is the index of the Landau level; this is in contrast to conventional two-dimensional electron systems such as GaAs and Si, which follow the selection rule $\Delta N =\pm 1$. Because of the unequal spacing between Landau levels and the unconventional optical selection rule, inter-Landau-level transitions (so-called cyclotron resonance) cover a broad THz frequency range that cannot be achieved with other materials. Therefore, graphene has been recently received considerable attention for use in THz emitter and photodetector applications [3-7].

Several studies have been conducted to detect inter-Landau-level transitions by means of photocurrent or photovoltage measurements [8-12]. Previously, the bolometric effect, photovoltaic effect, and microwave rectification have been considered to be the sources of the photo-induced signal in graphene. However, recent experiments demonstrated that the photo-thermoelectric effect has a more significant contribution to the photo-induced signal compared to other mechanisms owing to the large thermoelectric coefficient of graphene [13-18]. Thus far, this effect has not been seriously considered for detecting cyclotron resonance in graphene, primarily because of the symmetric carrier-density profile used in most of the previous devices, as we will discuss in a subsequent paragraph. In the present study, we fabricated a double-back-gated graphene device to create an asymmetric carrier-doping profile in the device and demonstrated the photo-thermoelectric detection of cyclotron resonance.

A schematic of the fabricated device and its optical micrograph are shown in Figs. 1(a) and 1(b), respectively. Graphene was encapsulated with two different layers of h-BN. The thicknesses of both the top and bottom h-BN layers are ~40 nm. Under the bottom h-BN layer, a graphite layer with a thickness of ~7 nm was placed as a local back-gate electrode. The device was fabricated on a 290-nm-thick $SiO_2$/p-doped Si substrate. By applying a gate voltage $V_R$ to the graphite gate, the carrier density $n_R$ of the right-side graphene located on the graphite gate is changed. By applying a gate voltage $V_L$ to the p-doped-Si substrate, the carrier density $n_L$ of the left-side graphene located on the p-doped Si substrate is changed (the electric field from the p-doped Si substrate is completely screened by graphite). This structure was fabricated using mechanical exfoliation from graphite and a bulk crystal of h-BN as well as the dry transfer of each layer using a method based on polymethyl methacrylate (PMMA) [19,20]. Using electron beam (EB) lithography and EB evaporation, a Pd contact electrode with a thickness of 120 nm was fabricated. Measurements were performed using a variable-temperature cryostat, and the temperature during the measurement was maintained at 3 K. A magnetic field was applied perpendicular to the plane from a superconducting magnet. A wavelength-tunable $CO_2$ laser ($\lambda$ = 9.28–10.61 μm) was used to irradiate the sample. By using an optical fiber and a polished stainless-steel light pipe, the laser light is delivered to the sample with a spot size of ~33 mm$^2$ and power density of 2.71 × 10$^{-2}$ Wcm$^{-2}$. The light was modulated using an optical chopper with a frequency of 85 Hz, and the photo-generated voltage was measured using a lock-in amplifier. The intensity of the irradiated light is homogeneous across the device because the light spot is much larger than the size of graphene. In addition, we assume that the irradiated light becomes non-polarized at the sample surface because of the use of the optical fiber and light pipe. To measure the two-terminal resistance of the graphene layer, a lock-in amplifier was used with an ac-current excitation amplitude of 10 nA.

First, we illustrate the generation of photo-thermoelectric voltage in a two-terminal graphene device having a symmetric carrier-doping profile in Fig. 1(c). The carrier density $n(x)$ is constant between the two Pd electrodes. Under homogeneous infrared light irradiation, graphene's electron temperature $T(x)$ increased significantly owing to its small electron heat capacity and small electron-phonon coupling. This creates a temperature difference between the graphene channel that is not covered with a metal contact and the graphene under the metal contact. Subsequently, a thermoelectric voltage $V(x) \propto S(n) \cdot (dT/dx)$ is generated in the vicinity of the junction between the graphene channel and the metal-covered graphene, where $S(n)$ denotes the Seebeck coefficient of graphene and depends on the carrier polarity as well as carrier density [13,14]. For the case of a uniformly doped device as shown in Fig. 1(c), the photo-thermoelectric voltages generated at the left and right graphene/metal junctions have the same amplitude but opposite signs. Since the measured photo-thermoelectric voltage is given by $V_{\text{induced}} = \int V(x) dx$, we obtain $V_{\text{induced}} = 0$ for the case of Fig. 1(c) owing to the cancellation between the $V(x)$ values of left-side and right-side graphene. This prevents the detection of the photo-thermoelectric effect in the two-terminal measurement. To overcome this issue, we used two different back gates $V_L$ and $V_R$, as shown in Fig. 1(d). The carrier densities of the left-side and right-side graphene, $n_L$ and $n_R$, respectively, can be controlled independently, enabling us to create asymmetric doping between them. Owing to this asymmetry, $S(n)$ also becomes asymmetric; consequently, the photo-thermoelectric voltages generated at the left-side and right-side graphene do not cancel each other. Particularly, we focus on the case in which a large $n_R$ is induced. As illustrated in Fig. 1(d), we can make $V(x)$ on the right-side graphene negligibly small when a large $n_R$ is induced because $S(n)$ scales with $1/\sqrt{n}$ [13-18]. We apply a large constant $V_R$ to keep a large $n_R$ value, while we measure $V_{induced}$ under the sweep of $V_L$. This enable us to detect the $n_L$ dependence of photo-thermoelectric voltage generated at the left-side graphene/meal contact junction. Similarly, the photo-thermoelectric voltage generated at the right-side graphene/meal contact junction can be detected by sweeping $V_R$ while applying a large $V_L$.

The two-terminal resistances of the graphene under the sweep of $V_L$ were measured at zero magnetic field, and the results are shown in Fig. 2(a). During this measurement, $V_R$ was fixed to maintain a constant $n_R$ value of $-2.0 \times 10^{12}$ cm$^{-2}$. The results show a clear ambipolar modulation of carrier concentration with a field-effect mobility of 60,000 cm$^2$V$^{-1}$s$^{-1}$. Next, the photovoltage signal $V_{induced}$ was measured as a function of $n_L$ under laser irradiation with a wavelength of $\lambda = 10.182$ μm at zero magnetic field, as shown in Fig. 2(b). When $n_L$ is swept from a large positive value to the negative direction, $V_{induced}$ is positive and increases toward the Dirac point. $V_{induced}$ exhibits an abrupt sign change across the Dirac point and becomes negative in the negative $n_L$ region. The amplitude of the signal decreases toward a large negative $n_L$ value. Similarly, the $n_R$ dependence of graphene's resistance and $V_{induced}$ are shown in Figs. 2(c) and 2(d), respectively, and were measured with a fixed $n_L$ value of $3.2 \times 10^{12}$ cm$^{-2}$. The mobility of the right-side graphene was determined as 45,000 cm$^2$V$^{-1}$s$^{-1}$. Here, the sign of $V_{induced}$ in Fig. 2(d) is opposite to that in Fig. 2(b) such that photovoltage is negative (positive) when the $n_R$ is positive (negative). The abrupt change in the sign of $V_{induced}$ at the Dirac point and the reduction in the $V_{induced}$ signal when the carrier density is swept away from the Dirac point are reminiscent of the photo-thermoelectric effect [13-16,18]. Thus, we believe that these results constitute evidence that the photo-thermoelectric effect is selectively detected in our double-back-gated graphene device. In this picture, a positive (negative) $V_{induced}$ should be detected when the left-side (right-side) graphene is electron-doped, which is consistent with Figs. 2(b) and 2(d).

Next, we discuss the photovoltage measurement under a magnetic field. The $V_{induced}$ data in Figs. 2(b) and 2(d) are measured with magnetic fields of 1 and 2 T, and the results are shown in Fig. 2(e) and 2(f), respectively. With increasing magnetic field, oscillatory changes of $V_{induced}$ with respect to the carrier density are superimposed on the zero-field data. These associate with the Landau quantization of graphene's density of states, suggesting that the photo-thermoelectric effect is also sensitive to the splitting of Landau levels in graphene. Hereafter, we focus on the $n_R$ dependence of the $V_{induced}$ signal and investigate photo-thermoelectric effect in the high-magnetic-field regime.

Fig. 3(a) shows the magnetic-field dependence of the photovoltage signal in the field range of 0 to 12 T. From the $n_R$ and magnetic-field dependences of graphene's two-terminal resistance (not shown), we assigned quantum Hall plateaus with filling factors of $\nu = \pm 10, \pm 6$, and $\pm 2$ by dashed lines in Figs. 3(a) and 3(c). The cross section of Fig. 3(a) along a fixed carrier concentration of $n_R = -1.1 \times 10^{11}$ cm$^{-2}$ is shown in Fig. 3(b). We observed the resonant enhancement of $V_{induced}$ at a magnetic field of ~9 T. The signal at resonance ($B = 8.9$ T, indicated by the black arrow in Fig. 3(a)) is presented in Fig. 3(c).

We attribute the signal enhancement at ~9 T to the cyclotron resonance absorption of graphene. Further, we illustrate the energy levels of graphene's electrons under Landau quantization in Fig. 3(d). The electron energy of graphene is expressed as $E = \text{sgn}(N)v_F\sqrt{2e\hbar|N|B}$, where $v_F$ denotes the Fermi velocity of graphene, $e$ the elementary charge, $\hbar$ the Planck constant, $B$ the magnetic field perpendicular to graphene, and $N$ the Landau level index. Energy levels from $N = 0$ to $N = \pm 4$ are plotted in Fig. 3(d). When the energy of irradiated light satisfies the selection rule of graphene ($\Delta|N| = \pm 1$), light absorption occurs. This is so-called cyclotron resonance. In comparison with Fig. 3(a), we found that the energy of irradiated light, 121.8 meV (corresponding to $\lambda = 10.182$ µm), is in good agreement with the transition from $N = 0 \rightarrow N = 1$ and from $N = -1 \rightarrow N = 0$ at ~9 T, as illustrated in Fig. 3(d); this is the lowest transition of the Landau-quantized graphene. In addition to the high field resonance, we also observed resonances in the low-magnetic-field region of Fig. 3(a), as shown in Figs. 3(e-g). Three resonance peaks at ~1.3, ~0.7, and ~0.5 T are observed, as indicated by red, green, and orange arrows, respectively, in Fig. 3(e). The cross section of Fig. 3(e) along a fixed carrier concentration of $n_R = 1.8 \times 10^{10}$ cm$^{-2}$ is shown in Fig. 3(f). The signal at resonance ($B = 1.3$ T, indicated by the red arrow in Fig. 3(e)) is presented in Fig. 3(g). The Landau energy diagram is shown in Fig. 3(h); in comparison with experimental data, the energy of irradiated light can satisfy transitions of $N = -1 \rightarrow N = 2$ and $N = -2 \rightarrow N = 1$ (2nd transition: red arrows), $N = -2 \rightarrow N = 3$ and $N = -3 \rightarrow N = 2$ (3rd transition: green arrows), and $N = -3 \rightarrow 4$ and $N = -4 \rightarrow 3$ (4th transition: orange arrows) at field values of ~1.3 T, ~0.7 T, and ~0.5 T, respectively. Thus, the appearance of three resonance peaks at low-field regions constitutes further evidence for the detection of cyclotron resonance in our experiment. Note that, in Fig. 3(a), the position of the cyclotron resonance peak shifts to a lower magnetic field around the Dirac point. A similar shift was reported by another group and was attributed to the change of graphene's effective Fermi velocity [21,22].

The changes of resonance magnetic field with different wavelengths of irradiated light are shown for both the 1st and 2nd transitions in Fig. 4(a). Here, the data are taken at the carrier concentration for maximum positive values of the photovoltage, which approximately corresponds to $\nu = -0.5$. As the wavelength increases, the magnetic field for cyclotron resonance shifts to a lower magnetic field. The peak position is summarized against the peak magnetic-field values in Fig. 4(b) for the 1st, 2nd, and 3rd resonances. By using the relation $E = \text{sgn}(N) v_F \sqrt{2e\hbar|N|B}$, we fit the data using the Fermi velocity $v_F$ as a fitting parameter, and the results are plotted as dashed lines in the figure. The obtained $v_F$ values are 1.1085, 1.1910, and $1.2063 \times 10^6$ m/s for the 1st, 2nd, and 3rd transitions, respectively. These values show good coincidence with the $v_F$ value given in the literature for similar h-BN-encapsulated graphene devices [21].

Finally, we discuss the shape of the $V_{\text{induced}}$ signal. The $V_{\text{induced}}$ signals at the 1st and 2nd cyclotron resonances shown in Figs. 3(c) and 3(g), respectively, increase in amplitude toward the Dirac point and reverse their signs at the Dirac point; this is similar to the case of zero magnetic field (Fig. 2). In addition, $V_{\text{induced}}$ exhibits an oscillatory change with respect to $n_R$, and the periodicity of oscillation matches a quantum Hall filling factor of $\Delta \nu = \pm 4$. Therefore, we believe that the $V_{\text{induced}}$ observed here is due to the magneto-photo-thermoelectric effect of graphene such that photo-induced heating is converted to voltage through the magneto-thermoelectric effect [23-25]. Under the magnetic field, the thermoelectric coefficient of graphene contains two different components: they are components longitudinal (Seebeck effect) and transverse (Nernst effect) to the thermal gradient [23-25]. Both thermoelectric coefficients approach zero at the quantum Hall plateaus such as $\nu = \pm 2, \pm 6$, and $\pm 10$, and they increase between the plateau regions. Therefore, the magneto-thermoelectric effect in graphene provides a $\Delta \nu = \pm 4$ sequence. Owing to the two-terminal geometry of our measurement, in principle, both the Seebeck and Nernst effect could contribute to the measured signal. To distinguish these effects, $V_{induced}$ was compared between different polarities of magnetic fields (supplementary Fig. S1). We found that the sign of $V_{induced}$ is unchanged upon reversal of the applied magnetic-field direction; thus, we believe that the main contribution is the Seebeck effect. Here, we estimated the temperature rise of graphene under light irradiation at the resonance condition ($B = 8.9$ T and $\lambda = 10.182$ μm) as ~2.8 K (supplementary Fig. S2). The Seebeck coefficient around this magnetic field and temperature is ~10 μV/K in graphene [23,24]. This value is in reasonably good agreement with our results for the photovoltage signal, showing a signal of ~80 μV with a temperature rise of ~2.8 K. Therefore, our result strongly suggests that the photo-thermoelectric detection of cyclotron resonance is achieved in graphene.

**Supplementary Material**

See supplementary material for the photo-induced voltage $V_{induced}$ measured with different polarities of magnetic fields and the temperature rise of graphene under light irradiation at the cyclotron resonance condition.

**Acknowledgements**

This work was supported by CREST, Japan Science and Technology Agency (JST) under Grant Number JPMJCR15F3, and by JSPS KAKENHI Grant Numbers JP25107001, JP25107003, JP25107004, JP26248061, JP15K21722, and JP16H00982.

**Figure captions**

Figure 1

(a) Schematic illustrations of the top and side views of the device fabricated in this study. (b) Optical micrograph of the device. (c,d) Schematic illustrations of the side view of the device, explaining the photo-thermoelectric detection scheme. These figures correspond to the cases of (c) a symmetric carrier-doping profile and (d) an asymmetric carrier-doping profile induced by tuning the right-side gate $V_R$.

Figure 2

(a) Two-terminal resistance of graphene $R$ and (b) photo-induced voltage $V_{induced}$ measured at 3 K under the sweep of $n_L$ at a fixed $n_R$ of $-2.0 \times 10^{12}$ cm$^{-2}$. (c) Two-terminal resistance of graphene $R$ and (d) photo-induced voltage $V_{induced}$ measured at 3 K under the sweep of $n_R$ at a fixed $n_L$ of $3.2 \times 10^{12}$ cm$^{-2}$. (e,f) Photo-induced voltage $V_{induced}$ measured at different magnetic fields of 0, 1, and 2 T. These are measured (e) under the sweep of $n_L$ at a fixed $n_R$ of $-2.0 \times 10^{12}$ cm$^{-2}$ and (f) under the sweep of $n_R$ at a fixed $n_L$ of $3.2 \times 10^{12}$ cm$^{-2}$.

Figure 3

(a,e) Magnetic-field and $n_R$ dependence of the photo-induced voltage $V_{induced}$ detected (a) from zero magnetic field to a high magnetic field and (e) in a low-magnetic-field region. (b,f) Cross sections of Figs. 3(a) and 3(e) along the magnetic-field direction at fixed carrier densities of (b) $n_R = -1.1 \times 10^{11}$ cm$^{-2}$ and (f) $n_R = 5.0 \times 10^{10}$ cm$^{-2}$, respectively. (c,g) The $V_{induced}$ signal under the sweep of $n_R$ measured at a resonance magnetic field $B$ of (c) $B = 8.9$ T indicated by the black arrow in Fig. 3(a) and (g) $B = 1.3$ T indicated by the red arrow in Fig. 3(e). Quantum Hall plateaus with the filling factor $\nu = \pm 2, \pm 6$, and $\pm 10$ are indicated by dashed lines in Figs. 3(a), 3(c), 3(e), and 3(g). (d,h) Energy diagrams of the electrons in graphene under the application of a perpendicular magnetic field. Coincident conditions of inter-Landau-level transition and optical excitation ($\lambda = 10.182$ μm) are depicted by arrows for (d) zero magnetic field to a high magnetic field (black arrow) and (h) a low-magnetic-field region (red, green, and orange arrows). All the data are obtained at a measurement temperature of 3 K and a fixed carrier density of $n_L = 3.2 \times 10^{12}$ cm$^{-2}$ in the left-side graphene.

Figure 4

(a) Magnetic-field dependence of the cyclotron resonance peak for the 1st resonance (right panel) and 2nd resonance (left panel) measured under different wavelengths of excitation. (b) Relationship between the energy of irradiated light and the magnetic field at which the cyclotron resonance occurred.

Figure 1

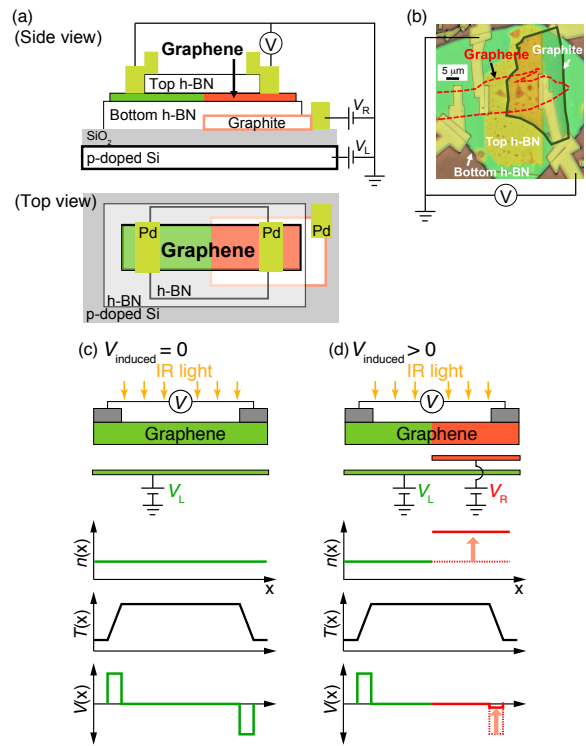

Figure 2

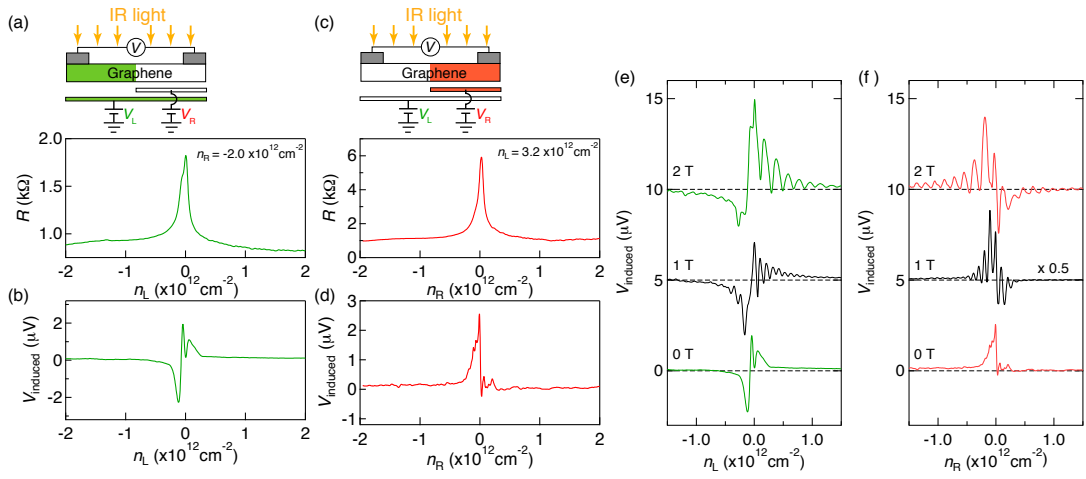

Figure 3

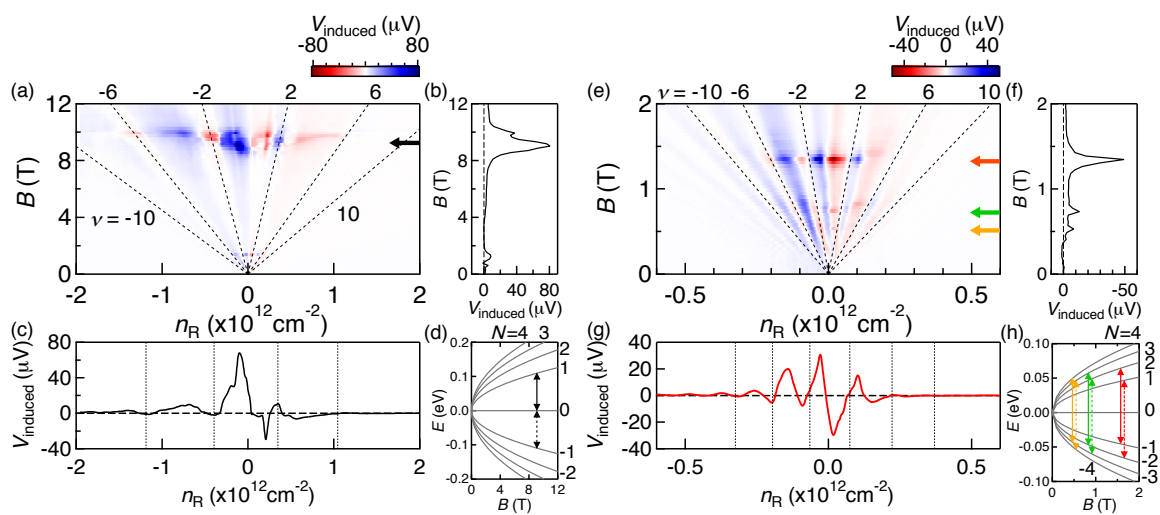

Figure 4

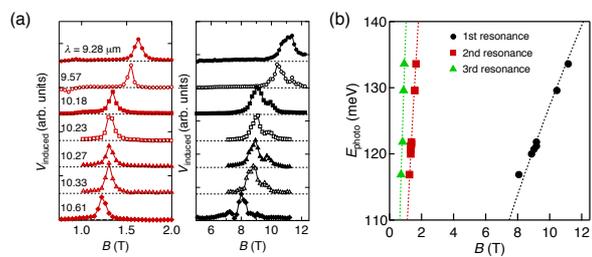